\documentclass{PoS}

\usepackage{epsfig}
\usepackage{multicol}

\title{Wilson Loop in Classical Lattice Gauge Theory and the Thermal Width of Heavy Quarkonium}

\ShortTitle{Wilson Loop in Classical Lattice Gauge Theory and the Thermal Width of Quarkonium}

\author{M.~Laine\\
Faculty of Physics, University of Bielefeld, D-33501 Bielefeld, Germany\\
        E-mail: \email{laine@physik.uni-bielefeld.de}}
	
\author{O.~Philipsen and \speaker{M.~Tassler}\\
Institute for Theoretical Physics, University of Münster, D-48149 Münster, Germany\\
        E-mail: \email{ophil@uni-muenster.de}, \email{marcus.tassler@uni-muenster.de}}

\abstract{\textbf{Abstract:} We present an estimate for the imaginary part of the recently 
introduced finite-temperature real-time static potential. It can be extracted from the time evolution of 
the Wilson loop in classical lattice gauge theory. The real-time static 
potential determines, through a Schr\"odinger-type equation and a subsequent
Fourier-transform of its solution, the spectral function of heavy quarkonium 
in finite-temperature QCD. We also compare the results of the classical
simulations with those of Hard Thermal Loop improved simulations, as well as 
with analytic expectations based on resummed perturbation theory.}

\FullConference{The XXV International Symposium on Lattice Field Theory\\
                 July 30-4 August 2007\\
                 Regensburg, Germany}

\renewcommand{\vec}[1]{\mathbf{#1}}
\newcommand{\rmi}[1]{{\mbox{\scriptsize #1}}}
\newcommand{\rmii}[1]{{\mbox{\tiny\rm{#1}}}}
\newcommand{\mD}{m_\rmii{D}}
\newcommand{\Nc}{N_{\rm c}}
\newcommand{\re}{\mathop{\mbox{Re}}}
\newcommand{\im}{\mathop{\mbox{Im}}}
\newcommand{\tr}{{\rm Tr\,}}

\newcommand{\figcaption}[1]{
  \begin{quote}
    \footnotesize{{\bf Figure} \arabic{figure}: #1}
  \end{quote}
  \vspace{0.1cm}
  \stepcounter{figure}
}

\newcounter{tabc}
\newcommand{\tabcaption}[1]{
  \vspace{-0.1cm}
  \begin{quote}
    \footnotesize{{\bf Table} \arabic{tabc}: #1}
  \end{quote}
  \vspace{0.1cm}
  \stepcounter{tabc}
}

\begin{document}

\section{Introduction}

Recently, we made an attempt to properly define a static potential in finite-temperature QCD, 
in the sense of obtaining an object which has a direct connection
to the spectral function of the heavy quarkonium system~\cite{static}.
The spectral function is related to a mesonic correlator obeying a Schr\"odinger equation
in real (Minkowski) time, and the corresponding potential was therefore introduced as the \emph{real-time static potential}. Furthermore, employing resummed perturbation theory, the real-time static potential 
was shown to develop an imaginary part, which induces a thermal width
for the tip of the quarkonium peak observed in the spectral function~\cite{og2}.
In a subsequent work~\cite{imV}, we investigated the extent to which
there might be non-perturbative corrections to the imaginary part, 
utilising classical real-time lattice techniques. The purpose of the
current note is to review the results of ref.~\cite{imV}, and also 
to elaborate on our Hard Thermal Loop improved simulations
in some more detail than in ref.~\cite{imV}.

\section{Real-time static potential}


The heavy quarkonium spectral function in the vector channel, $\rho(\omega)$, can be obtained using the relation
\begin{equation}\label{spectral}
 \rho(\omega) = \frac{1}{2} \Bigl( 1 - e^{-\frac{\omega}{T }}\Bigr)
 \int_{-\infty}^{\infty} \! {\rm d} t \, e^{i \omega t}
 C_{>}(t,\vec{0}) \;,
\end{equation}\vspace{-5mm}\\
where $C_{>}(t,\vec{0})$ is the mesonic correlator
\begin{equation}\label{correlator}
 C_{>}(t,\vec{r}) 
 \equiv 
 \int \! {\rm d}^3 \vec{x}\,
 \Bigl\langle
  \hat{\!\bar\psi}\,\Bigl(t,\vec{x}+\frac{\vec{r}}{2}\Bigr)
  \gamma^\mu
  \, W 
  \, 
  \hat \psi\Bigl(t,\vec{x}-\frac{\vec{r}}{2}\Bigr) \;\; 
  \hat{\!\bar\psi}\,(0,\vec{0})
  \gamma_\mu
  \hat{\psi}(0,\vec{0})
 \Bigr\rangle
 \;. 
\end{equation}
Here a point-splitting has been introduced to facilitate a perturbative treatment,  
and $W$ denotes a Wilson line connecting the adjacent operators along a straight path. 
The dilepton production rate from $q\bar{q}$-annihilation in a quark-gluon plasma 
is proportional to the thus defined spectral function.


Focusing on infinitely heavy quarks, 
the correlator can be obtained, up to normalization and a trivial phase factor, from 
the analytic continuation of a euclidean Wilson loop~\cite{static},
\begin{eqnarray}
 &&\hspace{3.5cm}C_>(t,\mathbf{r})\propto C_E(it,\mathbf{r}),\nonumber\\
 &&C_E(\tau,\mathbf{r})=\frac{1}{\Nc}\tr\left<W(0,\mathbf{r};\tau,\mathbf{r})W(\tau,\mathbf{r};\tau,0)
 W(\tau,0;0,0)W(0,0;0,\mathbf{r})\right> \;.
\end{eqnarray}
At $t\neq 0$ we can write the time evolution in the form of a Schr\"odinger equation,
\begin{equation}\label{Schroedinger}
 \left[i\partial_t-V_>(t,{r})\right]C_>(t,\vec{r})=0 \;, \quad r \equiv |\vec{r}| \;, 
\end{equation}
which defines the object $V_>$ we refer to as the \textit{real-time static potential}.


The simplest estimate for $V_>$ comes from perturbation theory. 
An analytic computation with proper account taken of HTL-resummation
yields the following result in the large-time limit~\cite{static}:
\begin{eqnarray}
&&
 V_{>}(\infty,r) = -\frac{g^2 C_F}{4\pi} \biggl[ 
 \mD + \frac{\exp(-\mD r)}{r}
 \biggr] - \frac{i g^2 T C_F}{4\pi} \, \phi(\mD r)\;, \nonumber\\
&&\textrm{with}\hspace{5mm}
 \phi(x) = 
 2 \int_0^\infty \! \frac{{\rm d} z \, z}{(z^2 +1)^2}
 \biggl[
   1 - \frac{\sin(z x)}{zx} 
 \biggr] \;.
\end{eqnarray}
The real part corresponds to the standard Debye-screened potential of a static quark--antiquark pair at finite temperature. 
The Debye mass is denoted by $\mD$. The imaginary part of the potential controls the damping of 
the correlator $C_>(t,\mathbf{r})$, which obeys the Schr\"odinger equation (\ref{Schroedinger}).


\begin{figure}[t]
\begin{center}
  \includegraphics[height=6.5cm,angle=0]{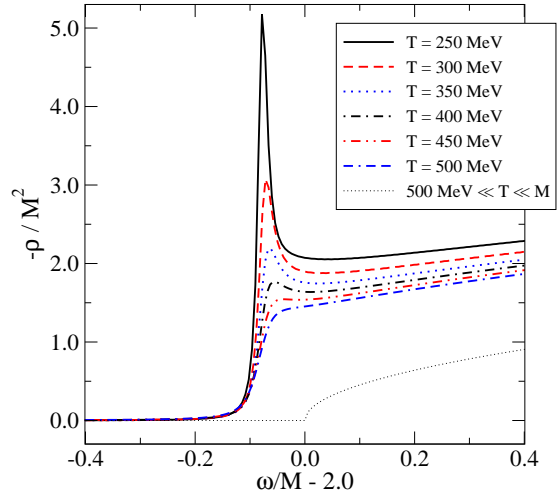}%
  \figcaption{The resummed perturbative quarkonium contribution 
  ($M$ denotes the heavy quark pole mass) 
  to the spectral function of the electromagnetic current,  
  in the non-relativistic regime $(\omega - 2 M)/M \ll 1$~\cite{og2}.}
\end{center}\vspace{-10mm}
\end{figure}

The spectral function
can now be obtained by inserting the static potential into the Schr\"odinger equation, 
eq.~(\ref{Schroedinger}), supplemented by the usual mass term and spatial derivatives, 
and employing subsequently eq.~(\ref{spectral}). 
The result is shown in fig.~1.
The imaginary part of the potential, 
encoding the Landau damping of the off-shell gluons binding the two heavy quarks together, 
introduces a thermal width to the tip of the quarkonium peak.

As a next step, we would like to estimate $V_{>}(\infty,r)$ beyond perturbation 
theory. (In principle, $\rho(\omega)$ could be extracted from lattice Monte
Carlo simulations by means of maximum-entropy and related methods; in practice, this involves many subtleties 
and, possibly, unknown systematic errors. For the current status see, e.g., refs.~\cite{ga}.) A non-perturbative calculation of $V_{>}(\infty,r)$ is complicated by the fact that 
a direct analytic continuation from numerical data for $C_E(\tau,\vec{r})$ is not feasible.  
It turns out, however, that the imaginary part of $V_{>}$ is formally classical~\cite{static}, 
and can hence be probed non-perturbatively with classical lattice gauge theory simulations, 
of the type originally introduced by Grigoriev and Rubakov~\cite{Grigoriev:1988bd}.

\section{Classical lattice gauge theory simulations}


We start our discussion of the real-time lattice techniques 
by introducing the framework for classical lattice gauge theory simulations~\cite{Grigoriev:1988bd}, 
which is quite similar to the Kogut-Susskind Hamiltonian approach~\cite{Kogut:1974ag}:
\begin{itemize}

\item 
The fields are discretized using a 3-dimensional spatial lattice. 
The time coordinate remains continuous.\vspace{-1mm}

\item Besides the spatial links $U_i$, corresponding to the discretized colour-magnetic fields, 
an electric field $E_i$ is defined via the relation $\dot{U}_i(x)=iE_i(x)U_i(x)$, 
where $x\equiv (t,\vec{x})$ and $\dot{U} \equiv \partial U/\partial t$.\vspace{-1mm}

\item A temporal gauge is chosen. The space of physical states is constrained to gauge field 
configurations satisfying the discretized Gauss law,
\begin{equation}
 G(x)\equiv \sum_i\left[E_i(x)-P_{-i}(x)E_i(x-\hat{i})\right]-j^0(x) \equiv 0  
 \;,
\end{equation}\\[-5mm]
with $j^µ$ denoting a possible colour current, 
and $P_i$ the adjoint parallel transporter, $P_i\phi(x+\hat{i})=U_i(x)\phi(x+\hat{i})U_i^\dagger(x)$.
\end{itemize}


The classical approximation for Yang-Mills fields at finite temperature follows 
by supplementing the phase space just introduced with a canonical time evolution 
and an average over initial conditions with a thermal weight. The weight corresponds 
to the one in the classical partition function,
\begin{equation}\label{partitionfunction}
 Z=\int \! \mathcal{D}U_i\, \mathcal{D}E_i\, \delta(G) 
 e^{-\beta H}\;,\hspace{5mm}
 H=\frac{1}{\Nc}\sum_x\biggl[ \sum_{i<j} \re\tr (1-U_{ij})+\frac{1}{2}\tr( E_i^2 )\biggr]
 \;,
\end{equation}
where $U_{ij}$ is the plaquette. 
The classical equations of motion for the discretized system can be obtained 
by invoking the Hamiltonian principle $\delta S=0$, 
and read [$\tr(T^a T^b) = \delta^{ab}/2$]~\cite{Ambjorn:1990wn}:
\begin{equation}\label{eom}
 \dot{U}_i(x)=iE_i(x)U_i(x) \;, \quad
 E_i = \sum_a E_i^a T^a \;, \quad
 \dot{E}_i^a(x)=- 2 \im \tr [T^a\sum_{|j|\neq i} U_{ij}(x)] 
 \;. 
\end{equation}


A more thorough treatment of the long-range dynamics of hot QCD is possible 
using the so-called Hard Thermal Loop (HTL) effective theory~\cite{Braaten:1989mz}, 
which is obtained by integrating out the ``hard modes'' (with momenta 
of the order of the temperature) from the system, in order to construct
an effective theory for the soft modes. To keep the effective theory local, 
certain on-shell particle degrees of freedom need, 
however, to be added to the effective Hamiltonian~\cite{Blaizot:1993zk}. 
Once this system is discretized and the classical limit is taken, 
the properties of the hard modes change, and the associated matching 
coefficient, denoted by $\mD^2$, needs to be tuned correspondingly~\cite{Bodeker:1995pp}. 
In the following we denote the new on-shell particle modes by $W(x,v)$. 
In a numerical implementation the following changes are introduced with respect to the 
classical setup:
\begin{enumerate}

\item
The Hamiltonian obtains an additional part,
\begin{equation}
 \delta H= \frac{1}{\Nc} \sum_x \biggl[ \int\frac{{\rm d}\Omega_v}{4\pi}\,  \frac{1}{2} (a \mD)^2 \tr(W^2) \biggr] 
 \;, 
\end{equation}
where $W\equiv T^aW^a(x,v)$ describes the charge density of the on-shell modes 
at $x$ moving in the direction $v=(1,\mathbf{v})$.

\item
The velocities $\vec{v}$ need to be discretised. This can be done, for instance, 
with spherical harmonics~\cite{Bodeker:1999gx} or with platonic solids~\cite{Rebhan:2005re}. 
Choosing the latter approach, we can replace $\int {\rm d}\Omega_v/4\pi f(v) \rightarrow 1/N_p \sum_{n=1}^{N_p} f(v_n)$, where $N_p$ is the number of vertices of the polyhedron used. 
The equation of motion of the gauge fields then acquires the source term
\begin{equation}
 j^{µ}(x)=(a\mD)^2 \frac{1}{N_p}\sum_{n=1}^{N_p} v_n^µW_n(x)
 \;, \quad W_n(x) \equiv W(x,v_n)
 \;.  
\end{equation}

\item
Finally, the new fields also evolve in time, according to the following equation of motion:
\begin{equation}
  \dot{W}_n(x)=v_n^i\left(\bar{E}_i(x)-\frac{1}{2} \left[P_iW_n(x+\hat{i})-P_{-i}W_n(x-\hat{i})\right]\right)
 \;,
\end{equation}
where $\bar{E}_i(x) \equiv [E_i(x)+ P_{-i} E_i(x-\hat{i})]/2$.
\end{enumerate}


For the purely classical simulations,
the required set of initial configurations distributed according 
to the statistical weight in eq.~(\ref{partitionfunction}) and respecting 
the Gauss constraint was created using the following algorithm:
\begin{enumerate} \vspace*{-1mm}
\item Pre-generate the spatial gauge links $U_i$ with a Monte Carlo simulation of the dimensionally reduced effective theory~\cite{Moore:1996qs}.\vspace*{-1mm}
\item Generate the electric fields from a gaussian distribution [cf.\ eq.~(\ref{partitionfunction})].\vspace*{-1mm}
\item Project onto the space of physical configurations, satisfying the Gauss law~\cite{Bodeker:1999gx}.\vspace*{-1mm}
\item Evolve the fields using the EOM, and repeat from step 2, until the fields have thermalized.\vspace*{-1mm}
\end{enumerate}
In the HTL-improved case there are minor changes, but the essence of the procedure is the same.
\vspace{-1mm}
\section{The imaginary part of the real-time static potential from Wilson loop dynamics}

\begin{figure}[t]
  \begin{center}
  \includegraphics[width=6.5cm,angle=0]{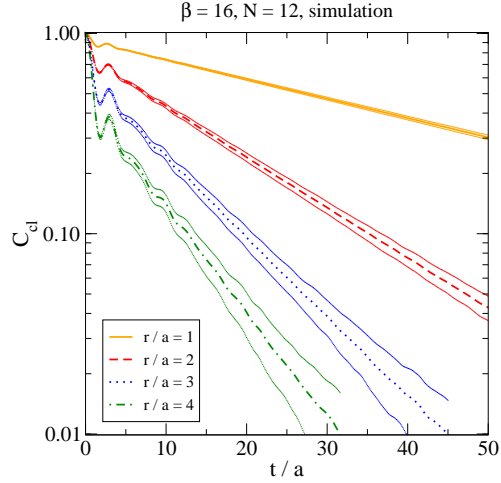}
  \figcaption{Measurement of the correlator $C_\rmi{cl}(t,r)$ [3]. The corresponding potential is shown in fig.~3.}
  \end{center}\vspace{-10mm}
\end{figure}

To obtain the imaginary part of the real-time static potential, a rectangular Wilson loop 
of spatial extent $r=|\vec{r}|$ and temporal extent $t$ was measured using classical 
or HTL-improved simulations. The measured average over a statistical
ensemble of initial configurations, as well as over lattice sites and loop orientations, 
is denoted by $C_\rmi{cl}(t,r)$ (a typical result is shown in fig.~2).
%
%
The real-time static potential can then be calculated from eq.~(\ref{Schroedinger}), 
\begin{equation}
 V_\rmi{cl}(t,r)\equiv \frac{i\partial_t C_\rmi{cl}(t,r)}{C_\rmi{cl}(t,r)}\;, 
 \hspace{5mm}C_\rmi{cl}(t,r)\equiv\frac{1}{\Nc}\tr\left<W^\dagger_r(t)W_r(0)\right>
 \;, 
\end{equation}
with $W_r(t)$ denoting a spatial Wilson line of length $r$. 
Timelike Wilson lines have disappeared due to the use of temporal gauge.
The result for $V_\rmi{cl}(t,r)$ is purely imaginary, 
and is shown in fig.~3.  

\begin{figure}[t]
  \begin{center}
    \includegraphics[width=6.5cm,angle=0]{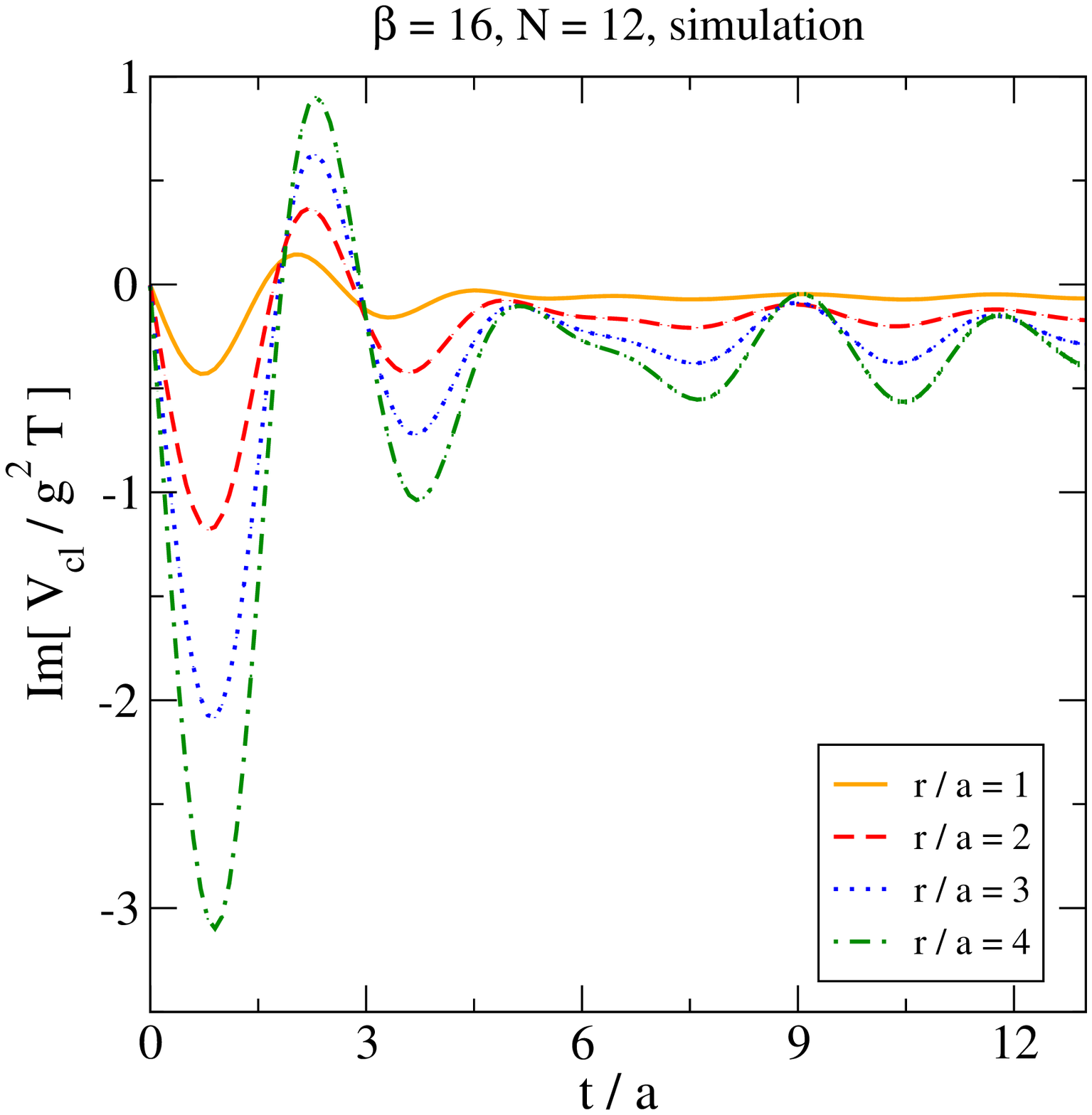}%
    ~~~~\includegraphics[width=6.5cm,angle=0]{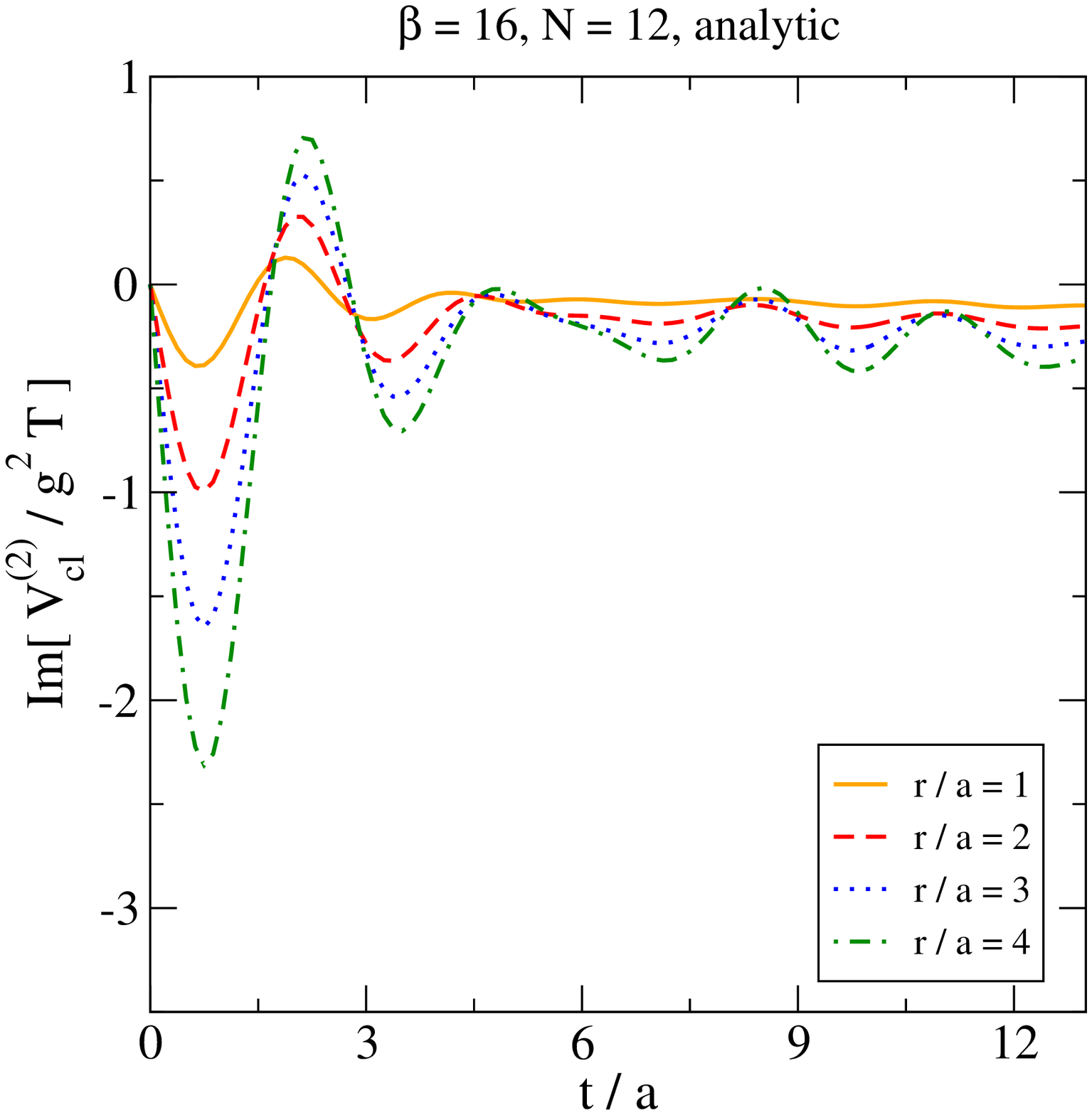}%
  \end{center}\vspace{-4.5mm}
  \figcaption{The imaginary part of the real-time static potential from the classical 
  simulation (left panel) and from resummed perturbation theory (right panel)~\cite{imV}.}
\end{figure}

\begin{table}[t]
\small\vspace{-1mm}
\begin{center}
\begin{tabular}{|c|c|c|c|c|c|c|c|c|}
\hline
&$\beta$&$N$&$a\mD$&confs&$r=1a$&$r=2a$&$r=3a$&$r=4a$\\
\hline
Simulation&16.0&12&0.0&200&-0.060(2)&-0.156(8)&-0.246(26)&-0.319(56)\\
&16.0&16&0.0&160&-0.059(2)&-0.155(8)&-0.245(22)&-0.326(48)\\
&16.0&12&0.211&200&-0.059(2)&-0.147(7)&-0.229(23)&-0.297(51)\\
\hline
\hline
&16.0&12&0.350&182&-0.030(2)&-0.064(5)&-0.096(12)&-0.118(21)\\
&13.5&12&0.250&142&-0.071(2)&-0.174(10)&-0.270(33)&-0.341(97)\\
\hline
\hline
Analytic&16.0&$\infty$&0.0&-&-0.0601&-0.1145&-0.1507&-0.1737\\
\hline
\end{tabular}\vspace{1.5mm}\\
 \tabcaption{Overview of the results in the large-time limit~\cite{imV}. 
 The results from the classical and HTL-improved simulations agree within error bars 
 for $a \mD < 0.25$ (at $\beta = 16$).}
\end{center}
\normalsize\vspace{-10mm}
\end{table}
As seen in fig.~3, the predictions from resummed perturbation theory 
and from the classical numerical simulations are remarkably similar. 
At the same time, some amplification of the imaginary part through the inclusion 
of non-perturbative (and higher-order perturbative) 
effects is visible in the simulation.
The difference between the two results becomes more pronounced at later times. 
In particular, in the large-time limit, a difference between the perturbative and 
the numerical results of up to $\sim 100\%$ can be observed (at $\beta = 16$), cf.\ table~1. 

\section{Conclusions}

The results from the real-time lattice simulations confirm 
the existence of an imaginary part in the real-time static potential, 
indicated already by leading-order Hard Thermal Loop resummed perturbation theory.
In fact, non-perturbative and higher order perturbative corrections {\em amplify} the imaginary part,
by up to $\sim 100\%$. The amplified imaginary part widens (and lowers) the quarkonium peak in fig.~1, 
although the qualitative structure remains unchanged. As a side remark, 
we note that the existence of an imaginary part also
leads to strong damping in the solution of the Schr\"odinger equation in eq.~(\ref{Schroedinger}), 
thus significantly facilitating the numerical determination of the spectral function
through eq.~(\ref{spectral}).
\newpage
\section*{Acknowledgements}

This work is part of the BMBF project \textit{Hot Nuclear Matter from Heavy Ion Collisions 
and its Understanding from QCD}.

\end{document}